# eXTP: enhanced X-ray Timing and Polarimetry Mission

S.N. Zhang[1], M. Feroci[2,58], A. Santangelo[1,3], Y.W. Dong[1], H. Feng[9], F.J. Lu[1], K. Nandra[15], Z.S. Wang[10], S. Zhang[1], E. Bozzo[4], S. Brandt[5], A. De Rosa[2], L.J. Gou[11], M. Hernanz[6], M. van der Klis[20], X.D. Li[13], Y. Liu[1], P. Orleanski[26], G. Pareschi[24], M. Pohl[7], J. Poutanen[42], J.L. Qu[1], S. Schanne[40], L. Stella[19], P. Uttley[20], A. Watts[20], R.X. Xu[16], W.F. Yu[12], J. J. M. in 't Zand[21], S. Zane[8], L. Alvarez[6], L. Amati[32], L. Baldini[36], C. Bambi[17], S. Basso[24], S. Bhattacharyya[47], R. Bellazzini[36], T. Belloni[24], P. Bellutti[59], S. Bianchi[38], A. Brez[36], M. Bursa[44], V. Burwitz[15], C. Budtz-Jorgensen[5], I. Caiazzo[28], R. Campana[32], X.L. Cao[1], P. Casella[19], C.Y. Chen[54], L. Chen[12], T.X. Chen[1], Y. Chen[1], Y. Chen[13], Y.P. Chen[1], M. Civitani[24], F. Coti Zelati[20,24,52], W. Cui[14,61], W.W. Cui[1], Z.G. Dai[13], E. Del Monte[2,58], D. De Martino[48], S. Di Cosimo[2], S. Diebold[3], M. Dovciak[44], I. Donnarumma[2], V. Doroshenko[3], P. Esposito[34], Y. Evangelista[2,58], Y. Favre[7], P. Friedrich[15], F. Fuschino[32], J.L. Galvez[6], Z.L. Gao[56], M. Y. Ge[1], O. Gevin[40], D. Goetz[40], D.W. Han[1], J. Heyl[28], J. Horak[44], W. Hu[1], F. Huang[54], Q.S. Huang[10], R. Hudec[44,45], D. Huppenkothen[37], G.L. Israel[19], A. Ingram[20], V. Karas[44], D. Karelin[6], P.A. Jenke[62], L. Ji[1], T. Kennedy[8], S. Korpela[41], D. Kunneriath[44], C. Labanti[32], G. Li[1], X. Li[1], Z.S. Li[29], E.W. Liang[58], O. Limousin[40], L. Lin[31], Z.X. Ling[11], H.B. Liu[58], H.W. Liu[1], Z. Liu[11], B. Lu[1], N. Lund[5], D. Lai[57], B. Luo[13], T. Luo[1], B. Ma[10], S. Mahmoodifar[63], M. Marisaldi[32], A. Martindale[8], N. Meidinger[15], Y.P. Men[59], M. Michalska[26], R. Mignani[34,35], M. Minuti[36], S. Motta[49], F. Muleri[2,58], J. Neilsen[51], M. Orlandini[32], A T. Pan[56], A. Patruno[22, 23], E. Perinati[3], A. Picciotto[59], C. Piemonte[59], M. Pinchera[36], A. Rachevski[43], M. Rapisarda[2,58], N. Rea[6,20], E.M.R. Rossi[22], A. Rubini[2,58], G. Sala[6], X.W. Shu[39], C. Sgro[36], Z.X. Shen[10], P. Soffitta[2], L.M. Song[1], G. Spandre[36], G. Stratta[50], T.E. Strohmayer[63], L. Sun[1], J. Svoboda[44], G. Tagliaferri[24], C. Tenzer[3], H. Tong[30], R. Taverna[33], G. Torok[46], R. Turolla[8,33], A. Vacchi[43], J. Wang[1], J.X. Wang[39], D. Walton[8], K. Wang[10], J.F. Wang[55], R.J. Wang[1], Y.F. Wang[56], S. S. Weng[53], J. Wilms[25], B. Winter[8], X. Wu[7], X.F. Wu[59], S.L. Xiong[1], Y.P. Xu[1], Y.Q. Xue[39], Z. Yan[12], S. Yang[1], X. Yang[11], Y.J. Yang[1], F. Yuan[12], W.M. Yuan[11], Y.F. Yuan[39], G. Zampa[43], N. Zampa[43], A. Zdziarski[27], C. Zhang[11], C.L. Zhang[1], L. Zhang[56], X. Zhang[13], Z. Zhang[10], W.D. Zhang[12], S.J. Zheng[1], P. Zhou[13], X. L. Zhou[11].

[1]Key Laboratory of Particle Astrophysics, Institute of High Energy Physics, Chinese Academy of Sciences, Beijing 100049, China; [2]IAPS-INAF, Via del Fosso del Cavaliere 100 - 00133 Rome, Italy; [3,1]IAAT University of Tuebingen, Sand 1 - 72076 Tuebingen, Germany; [4]ISDC, Geneve University, Chemin d'Ecogia 16 - 1290 Versoix, Switzerland; [5]National Space Institute, Technical University of Denmark, Elektrovej Bld 327, 2800 Kgs Lyngby, Denmark; [6]Institute of Space Sciences (IEEC-CSIC), Campus UAB, C/Magrans s/n, Barcelona, Spain; [7]DPNC, Geneve University, Quai Ernest-Ansermet 30 - 1205 Geneva, Switzerland; [9]Department of Engineering Physics and Center for Astrophysics, Tsinghua University, Beijing 100084, China; [10]Key Laboratory of Advanced Micro-structured Materials, Ministry of Education, Institute of Precision Optical Engineering, School of Physics Science and Engineering, Tongji University, Shanghai, 200090, China; [11]National Astronomical Observatories, Chinese Academy of Sciences, 20A Datun Road, Chaoyang District, Beijing, China; [12]Shanghai Astronomical Observatory, 80 Nandan Road, Shanghai 200030, China; [13]Nanjing University, 22 Hankou Road Nanjing Jiangsu 210093, China; [14]Purdue University, 525 Northwestern Avenue, West Lafayette, IN 47907, United States; [15]Max Planck Institute for extraterrestrial Physics, Giessenbachstr. 1, Garching, Germany; [16]Peking University, No.5 Yiheyuan Road Haidian District, 100871, Beijing, China; [17]Center for Field Theory and Particle Physics and Department of Physics, Fudan University, 200433 Shanghai, China; [18]GXU-NAOC Center for

Astrophysics and Space Sciences, Department of Physics, Guangxi University, Nanning 530004, China; [19]INAF-OA Roma, Via Frascati, 33 - 00040 Monte Porzio Catone, Italy; [20]Anton Pannekoek Institute, University of Amsterdam, Postbus 94249, Amsterdam, The Netherlands; [21]SRON, Sorbonnelaan 2 - 3584 CA Utrecht, The Netherlands; [22]Leiden Observatory, Niels Bohrweg 2 - NL-2333 CA Leiden, The Netherlands; [23]ASTRON, the Netherlands Institute for Radio Astronomy, Postbus 2, 7990 AA Dwingeloo, The Netherlands; [24]INAF - Brera Astronomical Observatory, via Bianchi 46, 23807, Merate (LC), Italy; [25]Remeis Observatory & ECAP, Universitaet Erlangen-Nuernberg, 96049 Bamberg, Germany; [26]Space Research Centre, Warsaw, Bartycka 18A - Warszawa, Poland; [27]Nicolaus Copernicus Astronomical Center, Polish Academy of Sciences, Bartycka 18 PL-00-716 Warszawa, Poland; [28]Department of Physics and Astronomy, University of British Columbia, 6224 Agricultural Road, Vancouver, BC V6T 1Z1, Canada; [29]Department of Physics, Xiangtan University, Xiangtan 411105, China; [30]Xinjiang Astronomical Observatory, Chinese Academy of Sciences, Urumqi, Xinjiang 830011, China;[31]Department of Astronomy, Beijing Normal University, Beijing 100875, China; [32]INAF-IASF Bologna, Via Gobetti 101, I-40129 Bologna, Italy; [33]Department of Physics and Astronomy, University of Padova, via Marzolo 8, 35131 Padova, Italy;[34]INAF-IASF Milano, via E.Bassini 15, I-20133 Milano, Italy; [35]Janusz Gil Institute of Astronomy, University of Zielona Góra, Lubuska 2, 65-265, Zielona Góra, Poland; [36]Istituto Nazionale di Fisica Nucleare, Sezione di Pisa, I-56127 Pisa, Italy; [37]Center for Data Science, New York University, 726 Broadway, 7th Floor, New York, NY 10003, USA; [38]University of Rome III, Via della Vasca Navale, 84 - 00146 Roma, Italy; [39]University of Science and Technology of China, No.96, JinZhai Road Baohe District, Hefei,Anhui, 230026, China; [40]CEA Saclay, DRF/IRFU, 91191 Gif sur Yvette, France; [41]University of Helsinki, Department of Physics, P.O.Box 48 FIN-00014 University of Helsinki, Finland; [42]Tuorla Observatory, Department of Physics and Astronomy, University of Turku, Väisäläntie 20, FI-21500 Piikkiö, Finland; [43]Istituto Nazionale di Fisica Nucleare, Sezione di Trieste, Via A. Valerio 2 - I-34127, Trieste, Italy; [44]Astronomical Institute of the Academy of Sciences of the Czech Republic, Fricova 298, CZ-251 65 Ondrejov, Czech Republic; [45]Czech Technical University in Prague, Zikova 1903/4, CZ-166 36 Praha 6, Czech Republic;[46]Silesian University in Opava, Na Rybníčku 626/1 - 746 01 Opava, Czech Republic;[47]Tata Institute of Fundamental Research, 1 Homi Bhabha Road, Colaba, Mumbai 400005, India; [48]INAF-OA Capodimonte, Salita Moiariello, 16 - 80131 Napoli, Italy; [49]Oxford University, Department of Physics, Clarendon Laboratory, Parks Road, Oxford, OX1 3PU, United Kingdom;[50]Università degli Studi di Urbino Carlo Bo, Piazza della Repubblica 13, I-61029, Urbino, Italy; [51]MIT, 77 Massachusetts Avenue - MA 02139 Cambridge, United States; [52]Università dell'Insubria Via Valleggio 11, I-22100 Como, Italy; [53]Department of Physics and Institute of Theoretical Physics, Nanjing Normal University, Nanjing 210023, China; [54]Shanghai Institute of Satellite Engineering, ShangHai 200240, China; [55]Department of Astronomy and Institute of Theoretical Physics and Astrophysics, Xiamen University, Xiamen, Fujian 361005, China; [56]Institute of Spacecraft System Engineering, Beijing 100094, China; [57]Cornell Center for Astrophysics and Planetary Science, Department of Astronomy, Cornell University, Ithaca, NY 14853, USA; [58]INFN, Sez. Roma Tor Vergata, Via della Ricerca Scientifica 1 - 00133 Rome, Italy; [59]Fondazione Bruno Kessler, via Sommarive 18, I-38123 Povo (Trento), Italy; [60]Purple Mountain Observatory, Chinese Academy of Sciences, Nanjing 210008, China; [61]Department of Engineering Physics and Center for Astrophysics, Tsinghua University, Beijing 100084, China; [62]University of Alabama in Huntsville, Huntsville, AL 35805, USA; [63]Astrophysics Science Division and Joint Space-Science Institute NASA Goddard Space Flight Center, Greenbelt, MD 20771, USA.

## ABSTRACT

eXTP is a science mission designed to study the state of matter under extreme conditions of density, gravity and magnetism. Primary goals are the determination of the equation of state of matter at supra-nuclear density, the measurement of QED effects in highly magnetized star, and the study of accretion in the strong-field regime of gravity. Primary targets include isolated and binary neutron stars, strong magnetic field systems like magnetars, and stellar-mass and supermassive black holes. The mission carries a unique and unprecedented suite of state-of-the-art scientific instruments enabling for the first time ever the simultaneous spectral-timing-polarimetry studies of cosmic sources in the energy range from 0.5-30 keV (and beyond). Key elements of the payload are: the Spectroscopic Focusing Array (SFA) - a set of 11 X-ray optics for a total effective area of ~0.9 $m^2$ and 0.6 $m^2$ at 2 keV and 6 keV respectively, equipped with Silicon Drift Detectors offering <180 eV spectral resolution; the Large Area Detector (LAD) - a deployable set of 640 Silicon Drift Detectors, for a total effective area of ~3.4 $m^2$, between 6 and 10 keV, and spectral resolution better than 250 eV; the Polarimetry Focusing Array (PFA) – a set of 2 X-ray telescope, for a total effective area of 250 $cm^2$ at 2 keV, equipped with imaging gas pixel photoelectric polarimeters; the Wide Field Monitor (WFM) - a set of 3 coded mask wide field units, equipped with position-sensitive Silicon Drift Detectors, each covering a 90 degrees x 90 degrees field of view. The eXTP international consortium includes major institutions of the Chinese Academy of Sciences and Universities in China, as well as major institutions in several European countries and the United States. The predecessor of eXTP, the XTP mission concept, has been selected and funded as one of the so-called background missions in the Strategic Priority Space Science Program of the Chinese Academy of Sciences since 2011. The strong European participation has significantly enhanced the scientific capabilities of eXTP. The planned launch date of the mission is earlier than 2025.

# 1. INTRODUCTION

In this paper we present the *enhanced X-ray Timing and Polarimetry* (eXTP) mission. This is an enhanced version of the XTP mission [1], which, in 2011, has been selected and funded for Phase 0/A as one of the background concept missions in the Strategic Priority Space Science Program of the Chinese Academy of Sciences. The XTP Phase 0/A study has been recently completed and successfully reviewed in May 2016. The scientific payload of eXTP consists of four main instruments: the Spectroscopic Focusing Array (SFA), the Polarimetry Focusing Array (PFA), the Large Area Detector (LAD) and the Wide Field Monitor (WFM). The fundamental scientific objective of the mission is the study of *matter under extreme conditions*, conditions not attainable in terrestrial laboratories. This fundamental goal will be reached by observing accreting neutron stars (NS) and black holes (BHs) in the X-rays, taking advantage of combining excellent timing capabilities, very good energy resolution, broadband spectral coverage and X-ray Polarimetry. The specific science objectives of the core program of the missions are: 1) The study of matter in ultra dense condition like the ones in the interior of neutron stars; 2) The physics and astrophysics of strong magnetic fields; 3) The study of accretion in strong-field gravity. An ambitious “observatory” program complements the core science program. eXTP is currently studied by an international consortium led by the Institute of High Energy Physics of the Chinese Academy of Science, and which includes many Chinese, European and US institutions, and among the latter a very large fraction of the LOFT collaboration [2, 3, 4, 5, 6]. The mission study is targeting launch earlier than 2025.

# 2. THE SCIENTIFIC PAYLOAD

A schematic view of the mission is shown in the left and right panels of Figure 1.

### 2.1 The Spectroscopic Focusing Array (SFA)

The SFA is an array of 11 identical X-ray telescopes covering the energy range 0.5-20 keV and featuring a total effective area larger than 0.6 $m^2$ at 6 keV, and ~0.9 $m^2$ between 1 and 2 keV (Figure 2, left panel). For each telescope, the requirement on the angular resolution is better than 1′ (HEW) while the Field of View (FoV) is expected to be about 12′ (FWHM). The telescopes are based on slumped glass optics (SGO) technology and are currently developed at the Institute of Precision Optical Engineering (IPOE), of the Tongji University, in China [7]. Different type of coatings -“no multilayer” and “multilayer”- are being studied to maximize the total area at 6 keV. Different solutions, which meet the requirements of 0.6 $m^2$ at 6 keV and 0.9 $m^2$ between 1 and 2 keV, have been obtained. In the current baseline the “no multilayer” option, consisting of only several layers of carbon, nickel and platinum or iridium, has been adopted. According to the current configuration the telescope focal length is 4.5 m for an aperture diameter of 450 mm.

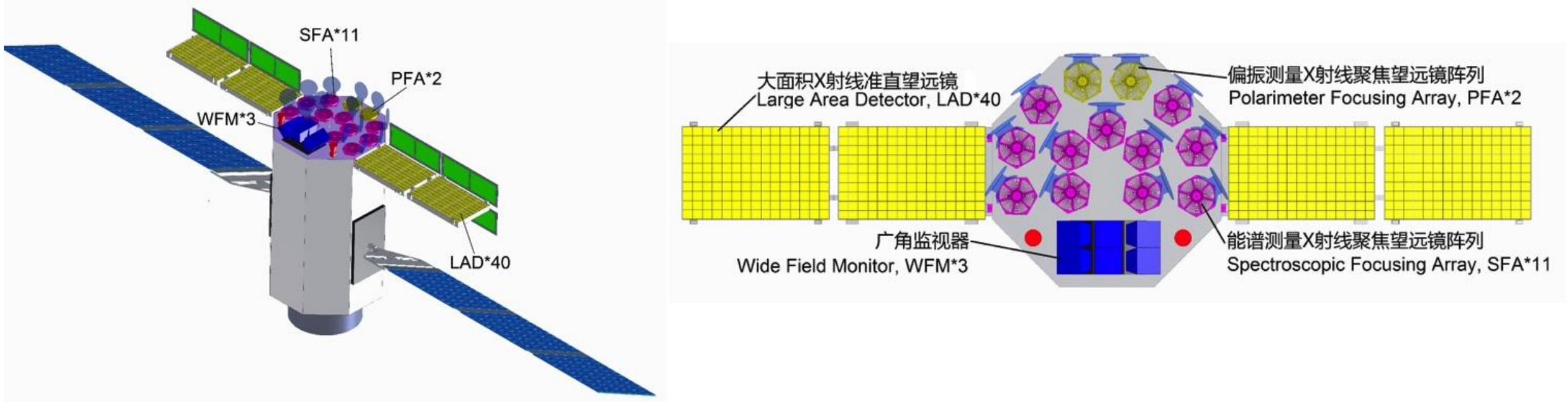

Figure 1. The baseline design of the mission is schematically summarized here. The left panel shows an artistic view of the eXTP satellite, with the payload module tower and four LAD panels. In the right panel the top view of the scientific payload is shown. The current configuration of the mission includes 11 telescopes of the SFA, 2 telescopes of the PFA, 40 modules for the LAD and 3 WFM units, each composed of 2 cameras.

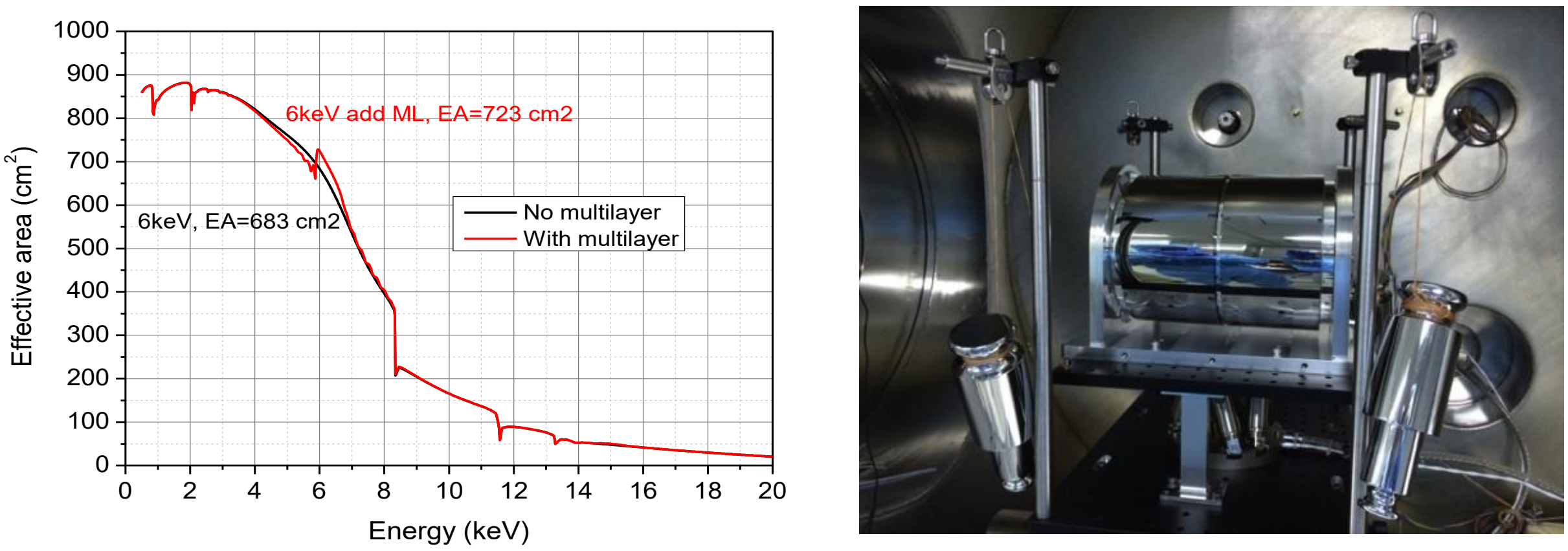

Figure 2. Left Panel: the effective area (single unit) of the current design of the eXTP SFA optics. On the right panel the very first prototype, successfully integrated at IPOE (Tongji University), is shown. The prototype has successfully passed vibration (5-6 g) and shock (5g) tests and its first performance testing session.

The baseline detector for the SFA is a 7 pixel Silicon Drift Detector (SDD). Studies on the SFA detector are led by the Max-Planck-Institut für extraterrestrische Physik, Garching (Germany). The pixel size is required to be smaller than 3′. The energy resolution (FWHM) will be better than 180 eV at 6 keV, while the time resolution of the instrument is 10μs. Currently a trade-off study is being performed to understand whether ASICs are a mandatory choice for the detector Front End Electronics or whether a solution based on discrete elements electronics can be implemented. A preliminary estimate of the SFA expected sensitivity is shown in Figure 3.

## 2.2 The Large Area Detector (LAD)

The LAD onboard eXTP uses the same design and technology of the LOFT mission. The eXTP LAD effective area,

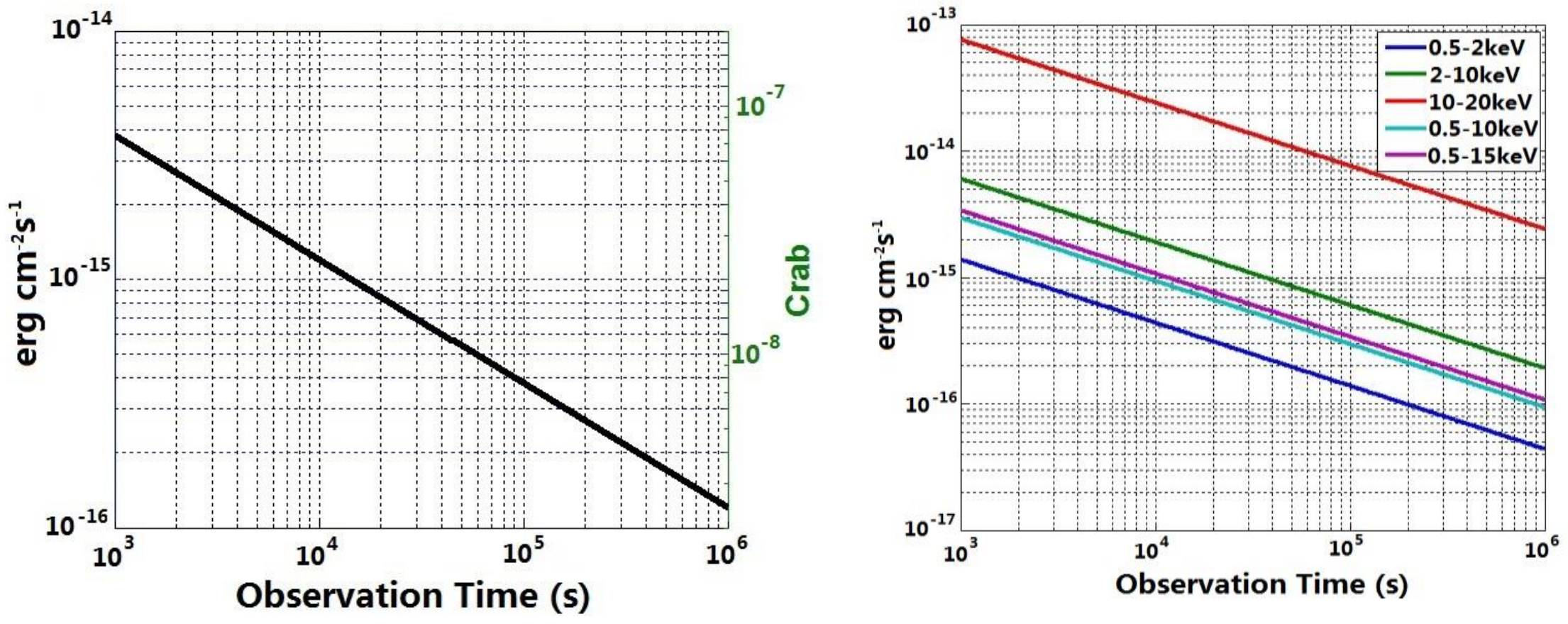

Figure 3. The sensitivity of the SFA in the 0.5-20 keV energy range (left panel) and for different energy ranges (right panel) is shown as a function of the exposure. These estimates are preliminary since studies on the background are still on-going

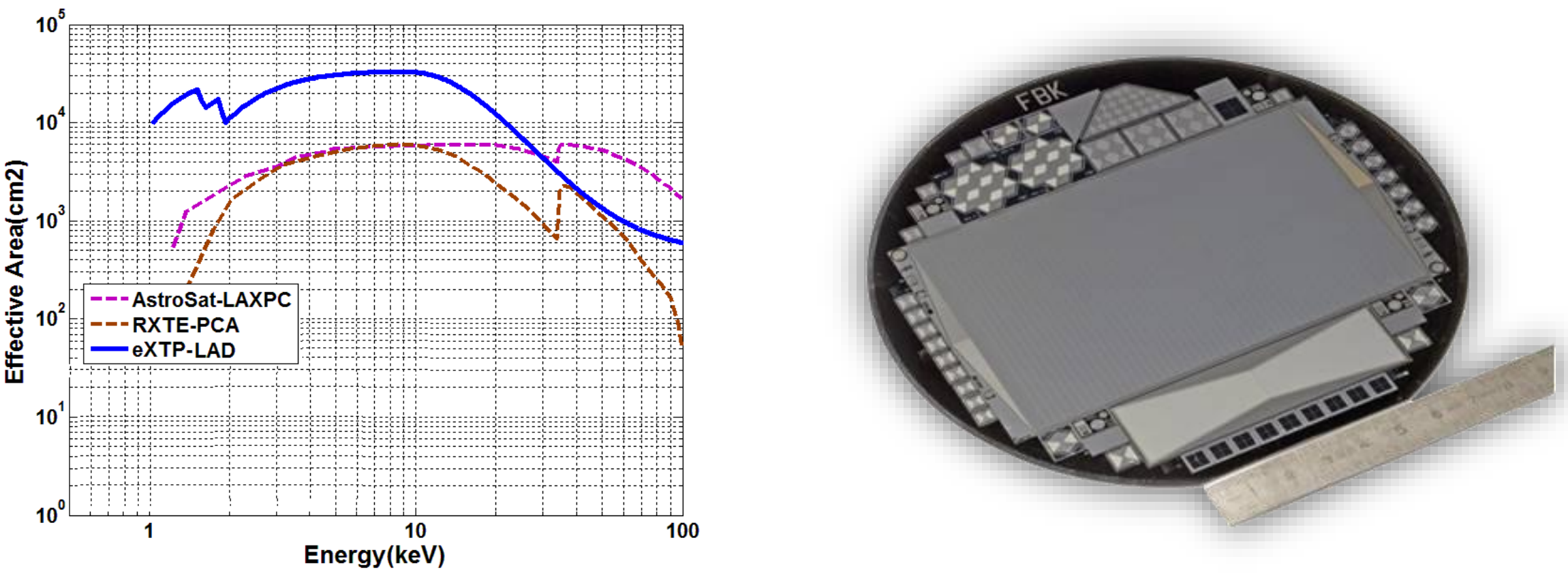

Figure 4. Left Panel: the eXTP LAD effective area is shown, on a log-log scale, in comparison with the one of the AstroSat-LAXPC and RXTE-PCA. It largely surpasses any past or currently flying mission and features an unprecedented value of about 3.4 $m^2$ at 6 keV. Right panel: an SDD manufactured by FBK (Italy). Several Prototypes have been manufactured and extensively tested in the context of the studies for the LOFT mission.

shown in Figure 4, reaches ~3.4 $m^2$ at 6 keV. The nominal energy range of the LAD is 2 -30 keV, but the dynamic range can be extended up to 80 keV to record bright events (e.g., GRBs or magnetar flares) shining through the collimator. The energy resolution is 250 eV at 6 keV. The FoV is limited to <1° FWHM by compact capillary plate (CP) collimators. The absolute time accuracy is 1μs while the dead time is less than 1% at 1 Crab flux level. The LAD reaches a sensitivity of 0.01mCrab for an exposure time of $10^4$s. The instrument consists of 40 modules of the same type developed for the LOFT mission. Each module in fact consists of a set of 4 x 4 detectors and 4 x 4 collimators, supported by two grid-like frames. The module hosts the read-out electronics, and the power supplies, organized in the Front-End Electronics (FEE) and Module Back-End Electronics (MBEE). The eXTP LAD modules are organized in large deployable panels, which host the Panel Back-end electronics (PBEE). The detectors are large-area SDDs originally developed for the ALICE/LHC experiment at CERN and optimized for the LOFT mission. The typical size is 11 x 7 $cm^2$ and 450 μm thickness [3]. Each detector is segmented in two halves with 112 channels each (970 μm pitch anodes). The rather compact capillary plate collimator is based on micro-channel plate technology. For the LAD, it consists of a 5-mm thick sheet of lead-glass (>40% Pb mass fraction) with the same size as the SDD detector, perforated by thousands of round micro-pores with 83μm diameter. The baseline CP collimator is based on Hamamatsu's round-pore technology, offering large open area ratio (75%) and a large Pb mass fraction. Alternative design options are currently being studied in China by Night Vision Technology Co, as well as in Europe, by Photonis (France). Ltd. More details on the LAD technology can be found in [6, 8, 10]. According to the current plans, the eXTP LAD will be procured by the institutions and countries already participating to the LOFT Consortium. The development of the instrument is led by IAPS-INAF (Italy) and MSSL (UK).

### 2.3 The Polarimetry Focusing array (PFA)

The Polarimetry Focusing Array consists of two identical telescopes with angular resolution better than 30″ (with a goal of 15″) and total effective area of about 250 $cm^2$ at 2 keV (see Figure 5). According to the actual baseline, optics are based on the well known Nickel technology, but a solution based on SGO is also being considered. The telescope features a focal length of 4.5 m with an aperture diameter of 450 mm. The Field of View is 12′. The focal assembly consists of two identical photoelectric X-ray polarimeters based on the Gas Pixel Detector concept (GPD, [12, 13]). The GPD is able to measure the linear polarization of photo-absorbed photons by reconstructing the emission direction of the ejected photoelectrons. The GPD comprises a gas cell with a thin 50 μm Beryllium entrance window, a Gas electron multiplier (GEM) and a pixelated charge collection plane, directly connected to the analog readout electronics. The GEM amplifies the charge of the electron tracks generated in the drift gap, without changing the track shape, and providing the energy and time information. Below the GEM, at less than a few hundreds micron, the top layer of the ASIC is covered by metal pads with a high filling factor distributed on a hexagonal pattern. Each pad is connected with an independent analog electronic channel. The ASIC, a development of INFN-Pisa, has 105600 pixels at 50 μm pitch, and it is at its third generation of development [14]. A prototype of the detector has been recently assembled and tested at Tsinghua University following a design by the INFN-Pisa group [11]. Thanks to an improved design a very careful manufacturing a good uniformity of the electric field has been obtained. The energy resolution (FWHM) is about 18% at 6 keV. The measured modulation factor very well agrees with the one predicted by simulations and reaches 0.6 above 6 keV. The systematic error for polarization measurement is less than 1% (at a confidence level of 99%). We refer to [11] for more details. The energy range of the PFA is 2-10 keV, and the time resolution is 500 μs. The sensitivity is about 5 μCrab for an exposure of $10^4$ s.

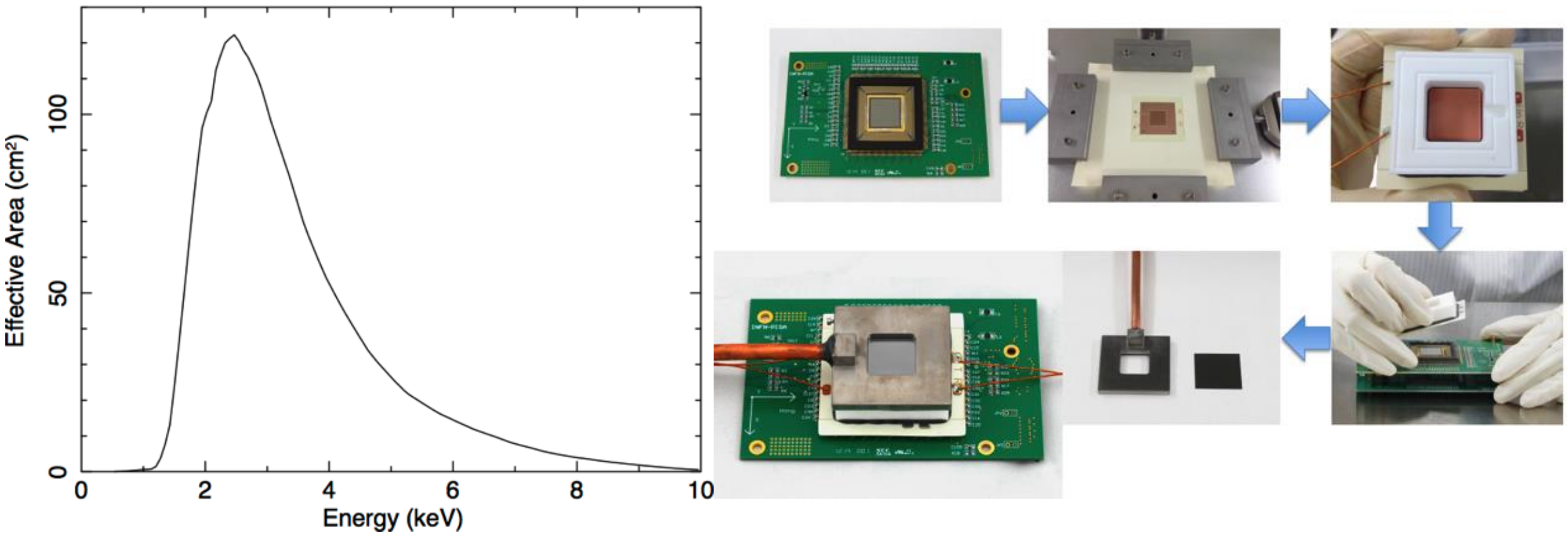

Figure 5. Left Panel: the single telescope PFA effective area. Right panel: different phases of the assembly of the GPD detector at the Department of Engineering Physics and Center for Astrophysics of the Tsinghua University (China). The composed picture has been adapted from [11]. The GPD prototype has been extensively tested and the performance is excellent.

According to the current baseline, the procurement of the PFA is led by the Chinese Team.

### 2.4 The Wide Field Monitor (WFM)

The WFM consists of three pairs of coded mask cameras covering 3.7 sr of the sky at a sensitivity of 4 mCrab for an exposure time of 1 d in the 2-50 keV energy range. The sensitivity, combining 1 yr of observations, reaches 0.2 mCrab outside the Galactic plane. The effective FoV of each camera pair is ~70° x70° (90° x90° at zero response). The energy resolution is ~300 eV at 6 keV, and the absolute time accuracy is 1 μs. The same SDDs as the LAD are implemented in the WFC but in a slightly modified geometry. In fact SDDs can provide accurate positions in one dimension and only rough position information along the second dimension. Therefore, pairs of two orthogonal cameras are combined to obtain rather precise 2D positions of the monitored sources. ASICs and Front End Electronics share a design similar to the one of the LAD-SDDs. To obtain the required position resolution, the WFC-SDD anode pitch is reduced to 145 μm (vs. 970 μm of the LAD SDDs). The required number of ASICs per SDD is higher (28x IDeF-X HD ASICs, with a smaller pitch than LAD, and 2x OWB-1 ASICs). The location accuracy is better than 1′, while the angular resolution is better than 5′. A 25 μm thick Beryllium window above each SDD protects against micrometeoroid impacts. The WFM uses the same BEE and PSU of the LAD, but with additional capability to determine photon positions. The ICU controls each of the six cameras independently and interfaces the PDHU, performing onboard computation to locate bright transient events in real time. More details on the WFM concept can be found in [6, 10].

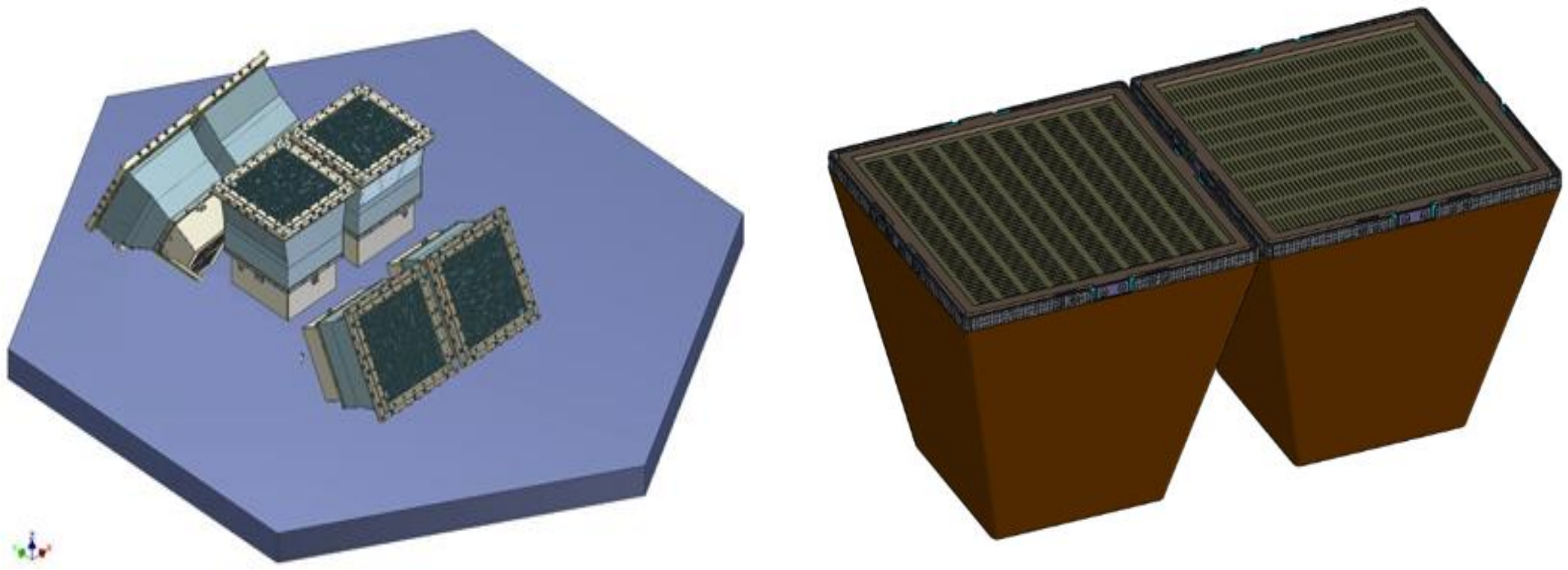

Figure 6. Left Panel: a schematic view of the three pairs of the WFM cameras. Right panel: cameras of each pairs are mounted orthogonally to obtain accurate 2D positions of the monitored sources.

## 3. THE SCIENCE CASE: MATTER UNDER EXTREME CONDITIONS

As we already mentioned the core science case of the mission is the study of the behavior of matter under extreme conditions that cannot be attained on earth. More specifically three key objectives constitute the main eXTP science program: the study of matter in ultra dense condition, the physics and astrophysics of strong magnetic fields, the physics of accretion in the strong-field limit of gravity. It is out of the scope of this paper to discuss in details the awesome and broad science capability of the mission. In what follows we will briefly summarize the key aspects of the mission science case and will present a few illustrative key examples. A series of four white papers is being finalized while this article is being published. The four white papers will focus on: i) *Dense Matter with eXTP*; ii) *Physics and Astrophysics of Strong Magnetic Field Systems with eXTP*; iii) *Accretion in Strong Field Gravity*; iv) *Observatory science with eXTP*. The awesome eXTP science case will be extensively discussed and argued in these white papers.

### 3.1 Dense matter with eXTP

One of the key goals of modern physics is to understand the nature of strong interactions, which determines the state of nuclear matter and sets the physics of neutron stars (NS), where gravity compresses matter to nuclear densities. Densities in NS cores can reach about 10 times the one of an atomic nucleus most likely forming “exotic” states and phases of matters, impossible to be realized in the laboratory: nuclear superfluids, strange matter such as hyperons and deconfined quarks, and the color superconductor phase [15]. As nicely summarized in Figure 7 (taken from [15]), observations of NS can allow accessing a unique regime of parameter space at high densities (e.g. high baryon chemical potential) and low temperatures, complementing therefore the research at the Large Hadron Collider and other heavy ion collision experiments, which aim to probe high temperatures and low densities. To connect strong interaction physics with observables we can use the NS equation of state (EOS) that relates pressure and density of the star. The EOS is encoded in the mass vs. radius diagram (M-R diagram) via the stellar structure equations. The knowledge of the M-R relation allows the determination of the NS EOS and enables the understanding of the microphysics at work in the extreme density regions in the interior of NS [15, 17]. The key observational step is the measurement of M and R with a few % precision and for several sources. Constraints obtained so far with different techniques, such as the modeling of the spectra of thermonuclear type I bursts or radio pulsar timing (see e.g. [15, 18, 19, 20] and references therein), are weak. eXTP will mainly use two techniques to constrain M and R for several NS: pulse profile modeling and spin measurements. Hotspots developing on a low magnetized, fast spinning NS give rise to observed pulse profiles that are strongly affected by GR light-bending and relativistic Doppler boosting and aberration. These relativistic effects, which depend on (for example) NS compactness M/R, strongly affect the amplitude of the pulsation and the asymmetry and harmonic content of the emerging profiles. By fitting high quality (i.e. high statistics) profiles M and R can be recovered in spite of degeneracies due to unknown factors like e.g. the geometry of the hotspot (size and inclination) and the observer inclination.

Figure 7. Left panel: The research for the physics of strong interactions spans a wide parameters space, ranging from

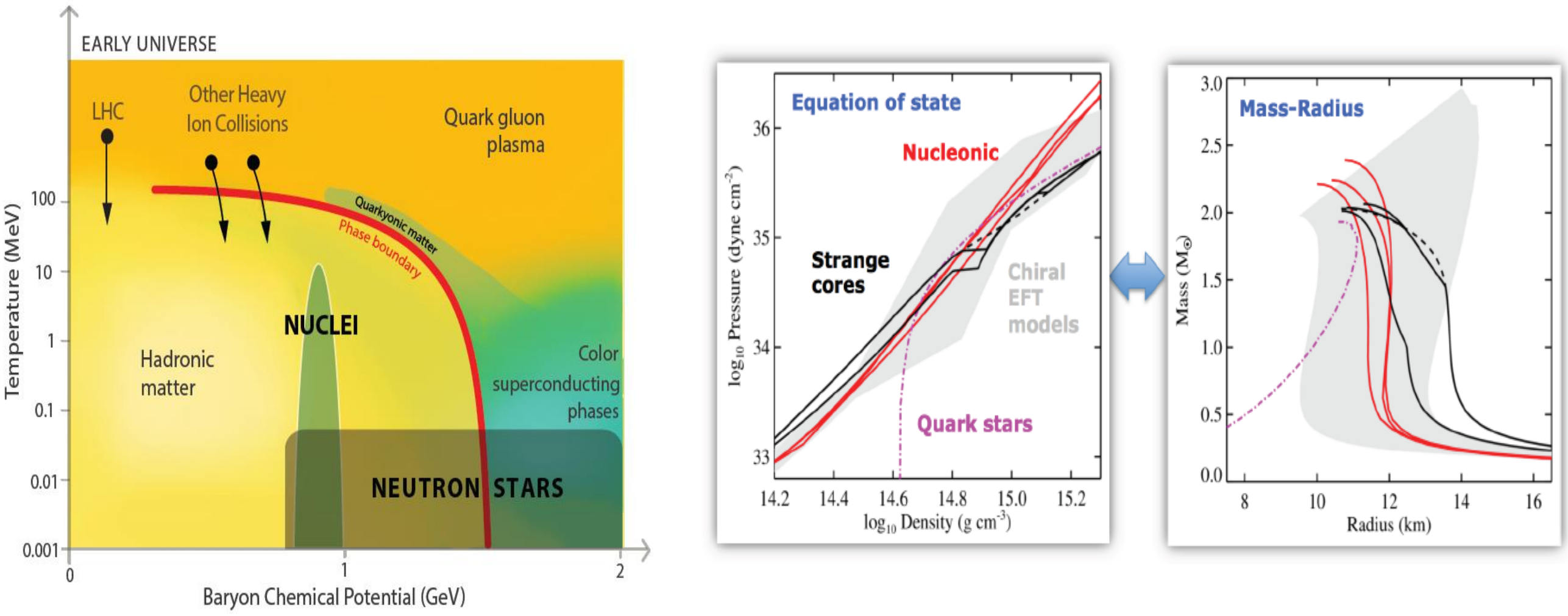

experiments at terrestrial accelerators, such as at LHC and RHIC, to observations of NS. In particular the low temperature vs. high Baryon chemical potential (e.g. extreme densities) regime can be explored by NS astrophysics. Progress in understanding NS matter will most likely reveal new matter states and phases. Figure is taken from the recent and exhaustive review by A. Watts and colleagues (2016) [15]. Right Panel: different EOS are predicted by different microphysics and convert to different M-R relations. More specifically the grey band corresponds to a range of nucleonic EOS based on chiral effective field theory. In red nucleonic EOS are shown. The black curves show the predictions of Hybrid models with strange quark core (black solid) and (black dashed) hyperon core model. Magenta: quark star model. The Figure has been taken and adapted from [16].

A unique feature of eXTP arises from its polarimetry capabilities. Radiation emerging from the hotspot on the NS surface is expected to be polarized and the observed polarization degree and angle are modulated with the spin phase. As discussed in detail in the seminal paper by Vironen and Poutanen (2004) [21], from phase resolved measurements of the polarization degree and angle both the observer and the hotspot inclination angles can be constrained largely reducing degeneracies. Accretion-powered millisecond pulsars (AMPs) and burst oscillation sources are ideal eXTP targets for pulse profile observations and modeling. In Figure 8 (courtesy of J. Poutanen) we present an example of the power of eXTP in combining the high statistics of the LAD with the PFA to constrain mass and radius of a bright AMP such as SAX J1808.4-3654, with a 100 ks observation. In general observations ranging from a few to several hundreds of ks are needed for each source. Although relatively large, these observing times are feasible and therefore within reach of the core program.

Constraints on the EOS can be obtained from he fastest spin rates and in particular more rapidly spinning NS place increasingly stringent constraints on the EOS (see [15, 23]). Since eXTP would have a larger effective area than any preceding X-ray timing mission, it is well suited to discover many more NS spins, using both burst oscillations and accretion-powered pulsations. One needs the large effective area of eXTP to detect burst oscillations in individual Type I X-ray bursts to amplitudes of 0.4 % (1.3%) rms in the burst tail (rise). As an example, preliminary simulations have shown that eXTP can perform a coherent search for intermittent pulsations down to amplitudes of 0.04 % rms (bright), 0.3% rms (moderate), 1.9% rms (faint).

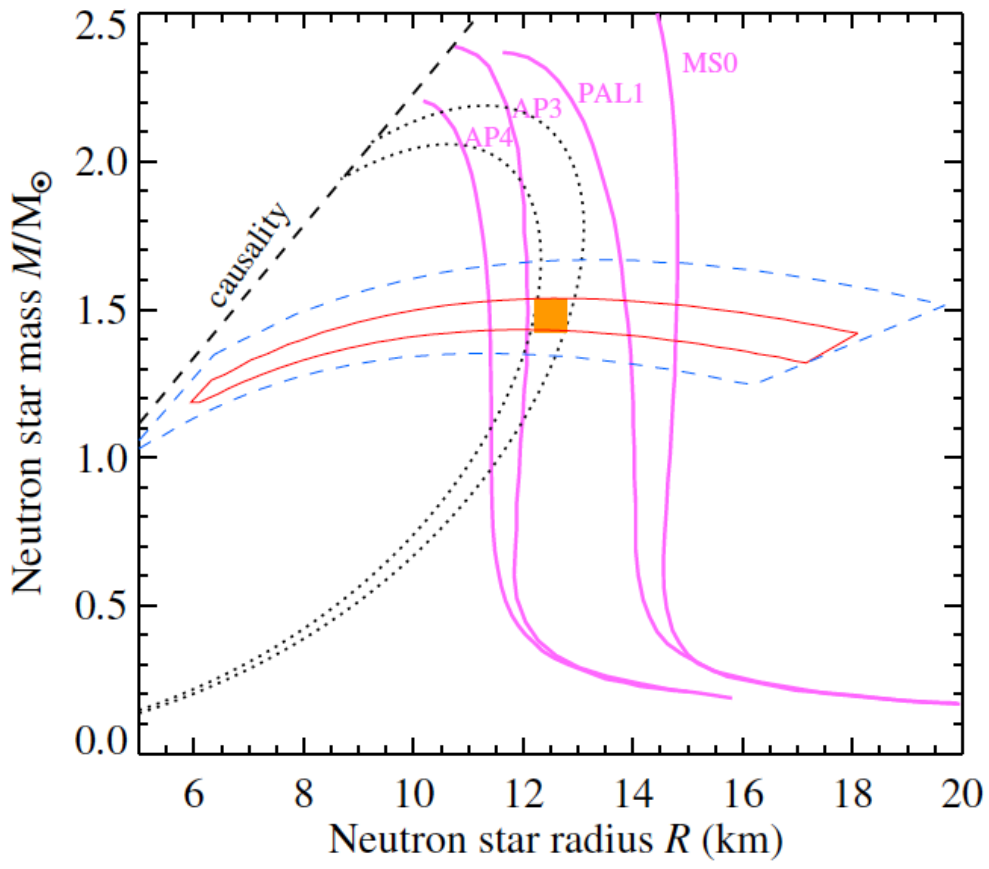

Figure 8. Constraints on M and R are shown as obtained from an eXTP 100 ks observation of SAX J1808.4-3658 compared to other existing observations. The blue dashed contours show M and R constraints from pulse profile modeling based on existing RXTE observations [22]. Similar constrains given by the eXTP LAD are shown by red solid contour and include the information obtained on the geometry with PFA observations. The nearly perpendicular dotted curves give constraints on M and R obtained by spectral evolution during photospheric radius expansion bursts as determined using the cooling tail method [18]. The pink solid curves correspond to different equations of state of cold dense matter. The overlapping region in orange gives a robust estimate of M and R at a few % level *(Courtesy of the eXTP working group on Dense Matter).*

### 3.2 Physics and Astrophysics of Strong Magnetic fields

The broadband, high sensitivity, polarimetry and monitoring capability of eXTP, can provide a deep understanding of the physics in extremely strong magnetic fields. Magnetars, accreting X-ray pulsars, and rotation-powered pulsars are key targets for eXTP. The mission will also enable observational studies of QED effects. Magnetars are highly variable X-ray sources, with magnetic fields of the order of $10^{14-15}$ G. Their flux can change by orders of magnitude. Variability occurs on different time scales, e.g. short bursts, (< 1s), intermediate flares (~1-40 seconds), giant flares (~500 s). Outbursts can last several weeks to years. Considering their variability and the WFM sky coverage, eXTP is expected to discover a new magnetar candidate every year, triggering follow up observations with the SFA and the LAD.

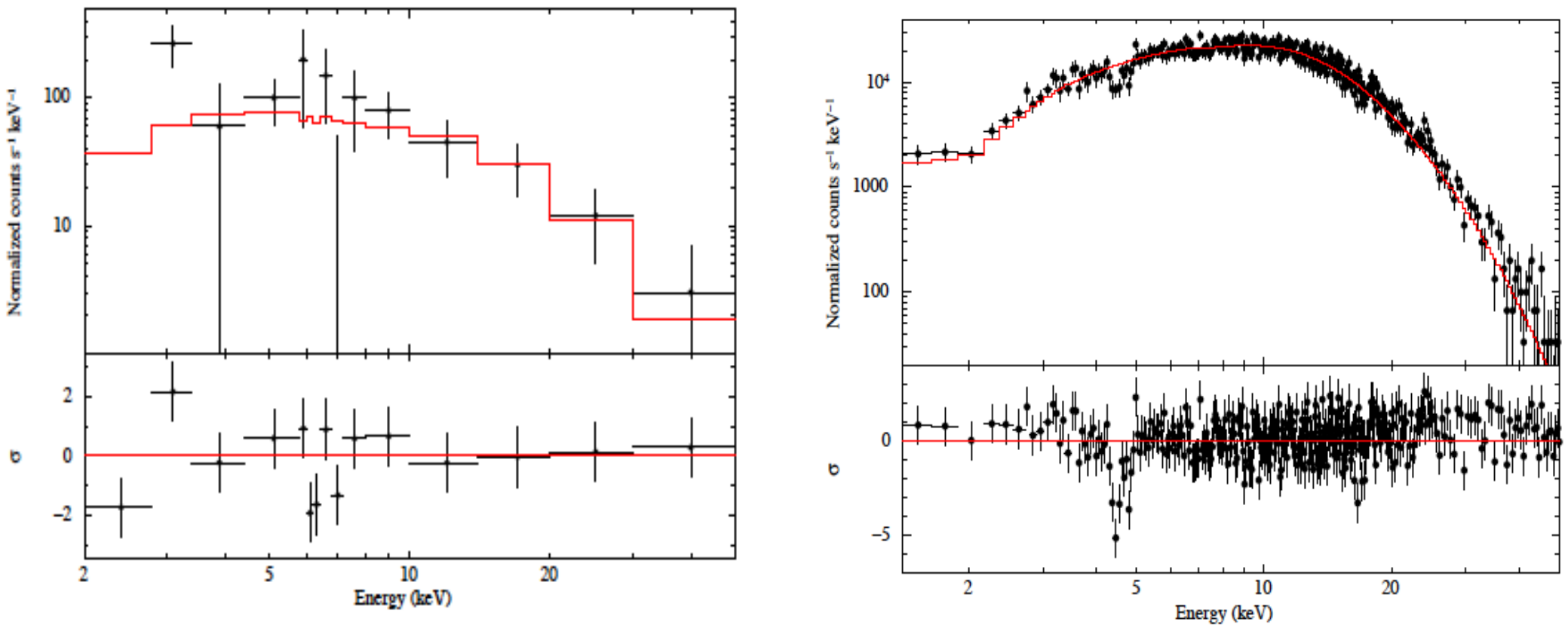

Figure 9. Left: simulated WFM spectrum of a 10 Crab magnetar burst (black points) lasting for 0.05 s. The assumed spectrum is a double blackbody. Right: Simulated LAD spectrum of the same burst, including an absorption feature at 4.5 keV. The lower panel shows the residual with respect to the best fit continuum and illustrates how the potential proton cyclotron line would be significantly observed. *(Courtesy of the eXTP working group on Strong Magnetism).*

As an example, in Figure 9 we illustrate how even a short duration (0.05 s) faint burst (10 Crab) is detected by the WFM with high significance. For the same burst a resonant scattering feature at 4.5 keV can be detected with high significance (>8σ). Thanks to the large field of view of the WFM a large fraction of the Galactic Plane will be covered during most eXTP pointings. This allows to monitor the spin period of several magnetars, obtaining phase-connected timing solutions. Through this monitoring, eXTP can detect glitches [24, 25], precession [26, 27], and will accurately measure braking indices [28, 29]. As shown in Figure 10, systematic time monitoring with eXTP will allow us to detect the amplitude of free precession down to a level of Δ f~10^{-9} Hz in a magnetar like SGR 1900+14.

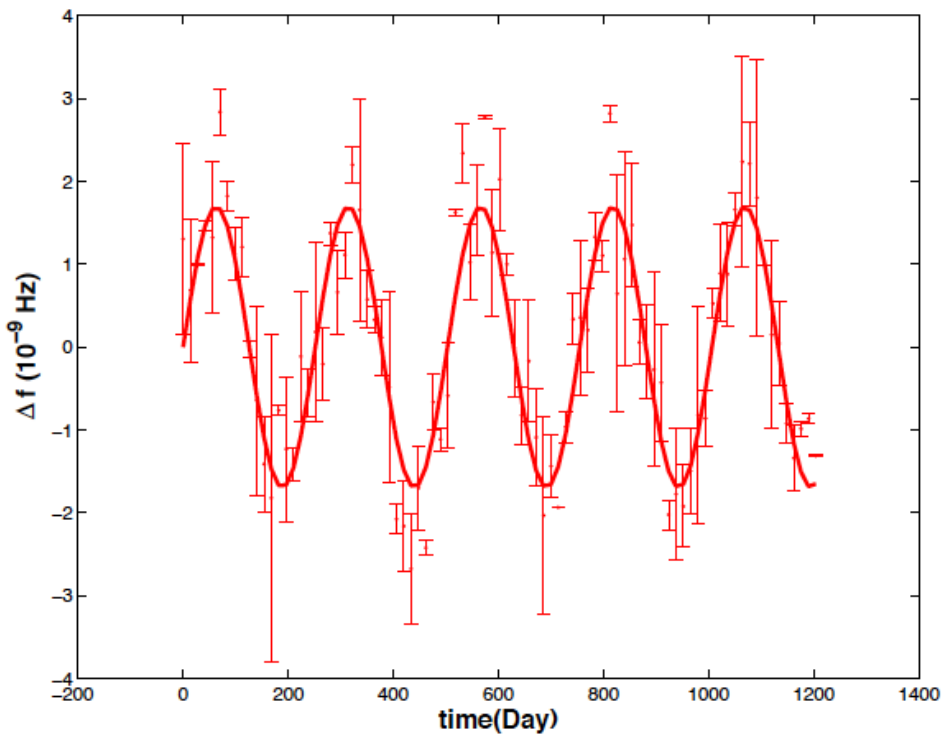

Figure 10. eXTP simulation of frequency variation due to free precession in SGR 1900+14. The input spectrum is an absorbed blackbody plus a power law. The hydrogen column density, the black body temperature and its normalization are the same as for SGR 1900+14, i.e. $1.6\times10^{22}$ $cm^{-2}$, 0.5 keV and 7.1 respectively. A sinusoidal modulation of the free precession is assumed. Each point has a duration of 2 ks, that for 2 weeks monitoring gives a total exposure time of $1.15\times10^{5}$ s. *(Courtesy of the eXTP working group on Strong Magnetism).*

Accreting X-ray pulsars are key targets for eXTP. Particularly relevant will be the observational studies on the polarization properties of the radiation emerging from these objects. In their seminal paper Meszaros et al. (1988) [30] have shown that the X-ray linear polarization depends strongly on the geometry of the emission region, and varies with energy and pulse phase, reaching very high degrees, up to 70% (see Figure 1 of [30]). We have estimated for several sources the exposure required to constrain the linear polarization fraction to better than 10% accuracy, which implies also that polarization angles will be significantly constrained too (Figure 11). For bright persistent sources, exposures of ~10 ks will be sufficient for phase resolved studies. For weak sources, significant polarization will be detected with exposures of ~100 ks, which still allows phase resolved studies with Ms exposures. The phase dependence of the polarization properties is closely related to the geometry of the emission region. This can be probed by eXTP through the analysis of the energy and luminosity dependence of pulse profiles and cyclotron lines. The long observations required for polarization studies will in addition yield high quality pulse profiles and will help to extend studies on the luminosity dependence of the cyclotron resonance scattering feature (CRSF) to low luminosity, probing the regime switch expected for accreting pulsars from local super-Eddington to sub-Eddington accretion (Figure 11) [31, 32, 33].

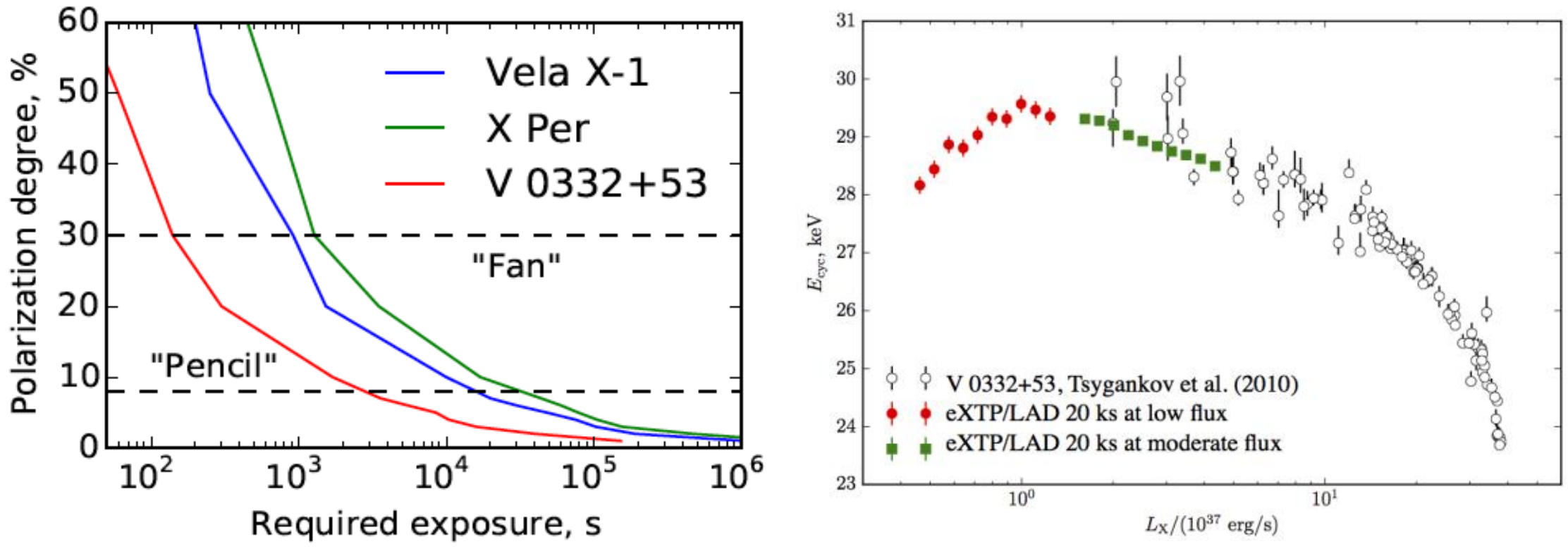

Figure 11. Left panel: We show the eXTP exposure required to constrain the linear polarization degree to better than 10% accuracy. With this exposure, polarization angles will be significantly constrained as well. Three cases are shown: 1) bright source (Vela X-1, $F_{2\text{-}10keV}$~$1.6\times10^{-9}$ erg $s^{-1}$ $cm^{-2}$); 2) a weak persistent source (X-Per, $F_{2\text{-}10keV}$~$6\text{x}10^{-10}$ erg $s^{-1}$ $cm^{-2}$); 3) a Be Transient in outburst (V 0332+53, $F_{2\text{-}10keV}$~$1.6\times10^{-8}$ erg $s^{-1}$ $cm^{-2}$). Right panel: The variation of the CRSF energy with luminosity is shown for V0332+53. LAD observations will allow accurate measurements of the line energy even at very low luminosities, extending the study of the correlation between the line and the luminosity on a very wide luminosity range. *(Courtesy of the eXTP working group on Strong Magnetism).*

The eXTP mission will provide the first tests of one of first predictions of QED: vacuum polarization and the effect of strong magnetic fields on the propagation of light [34, 35, 36]. Observations of NS can verify that this effect exists [37, 38, 39]. For magnetars this effect is the strongest, given their fields of ~$10^{14\text{-}15}$ G. Vacuum birefringence increases the expected linear polarization of the emerging X-rays from about 5-10% to nearly 100% [39]. It is nearly as strong for neutron stars with magnetic fields of $10^{12}$ G. An example of how eXTP observations will reach this goal is shown in Figure 12 where the observed pulsed light curve, and phase resolved degree and angle of polarization are shown for a 1 Ms observation of AXP 1RXSJ170849.0-400910 after properly accounting for QED effects.

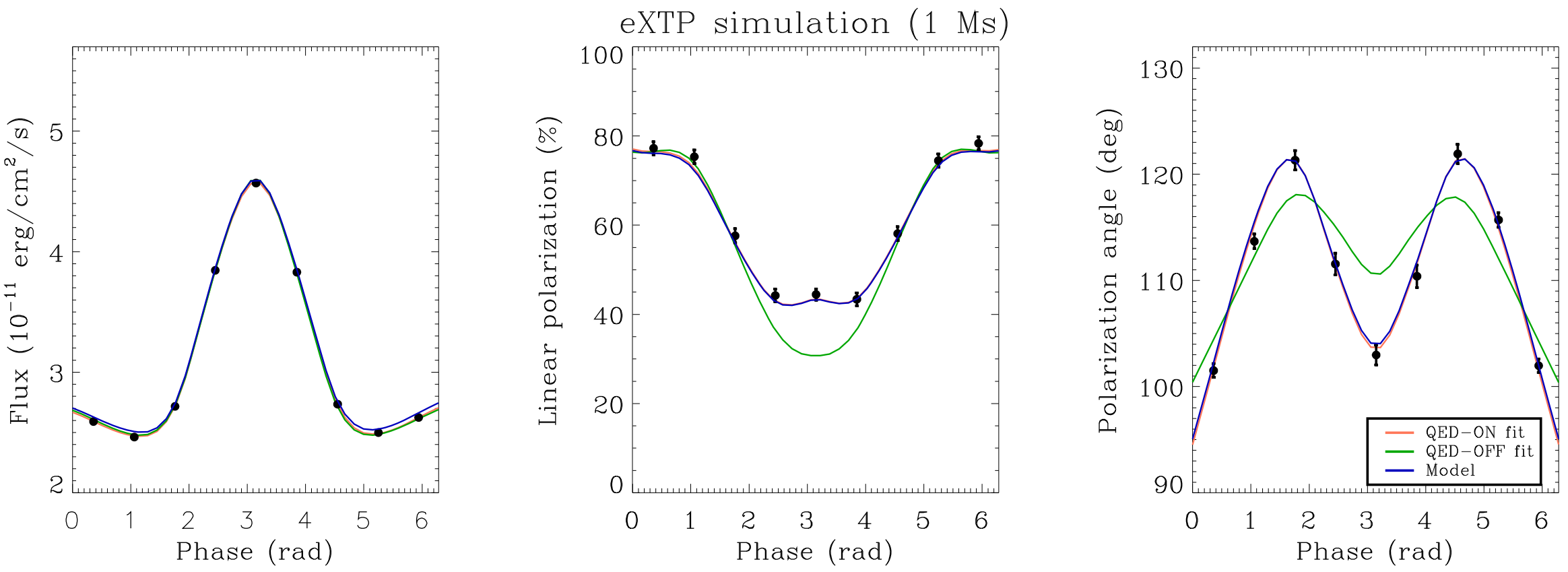

Figure 12. Pulse profile, and phase dependent degree and angle of polarization expected in 1Ms observation of AXP 1RXSJ170849.0-400910 assuming the twisted magnetosphere model [see 37 for details]. The points have been simulated taking into account QED effects (blue line). The red curve (no QED effects) is not consistent with the observed data. *(Courtesy of the eXTP working group on Strong Magnetism).*

### 3.3 Accretion in strong field gravity

One of the major challenges of modern astrophysics is the study of matter close to the event horizon of black holes, where gravity is in the strong-field regime. The motion of matter near super-massive black holes (in AGN) and stellar-mass black holes (in X-ray binaries), dominated by gravity, provides a powerful diagnostic to study the deep potential well generated by the central object, infer its mass and spin, and verify some of the crucial predictions of General Relativity (GR) in the strong-field regime. The two most important direct diagnostics of matter behavior in the strong-field regime are relativistically broadened Fe lines [40, 41] and relativistic time-scale variability, in particular quasi-periodic oscillations [QPOs, 42, 43, 44, 45]. eXTP combines the spectral resolution required for resolving relativistic lines with the large photon throughput required to study their variability on time scales down to well below the dynamical time scale of the strong-field region. In addition, polarization information will allow us, for the first time, to investigate the accretion/ejection flows around black hole using different techniques that will complement each other. In this paper we cannot fully present the eXTP capability to explore physical phenomena in the strong gravity regime.

We will, as before, focus on a few selected examples. In the simulation illustrated in Fig. 13, we show on the left panels the spectrum of a 0.5 Crab BH with maximal spin (e.g. GRO J1655-40, [46]). An eXTP observation can provide a measurement the inner radius of the disc and the radial emissivity with 1-2% accuracy in only 100s. Such unprecedentedly short timescale enables for the first time the study of the variability of the innermost region on a time-scale comparable to variations of outflow components such as winds and jets. This will open a new observational window on how the ejection properties are linked to the inner accretion flow. In the right panels of Figure 13, we show the eXTP spectrum obtained by a 100 ks integration of a 2 mCrab AGN, with spin parameter a=0.5. Our simulations show that in AGN the energy resolution of eXTP together with the large effective area and broadband energy coverage provided by the SFA and LAD combination, allow us to disentangle the spectral complexities in the Fe K region and measure the reflection continuum shape, to successfully extract the relativistic reflection parameters and recover the black hole spin to 10% accuracy.

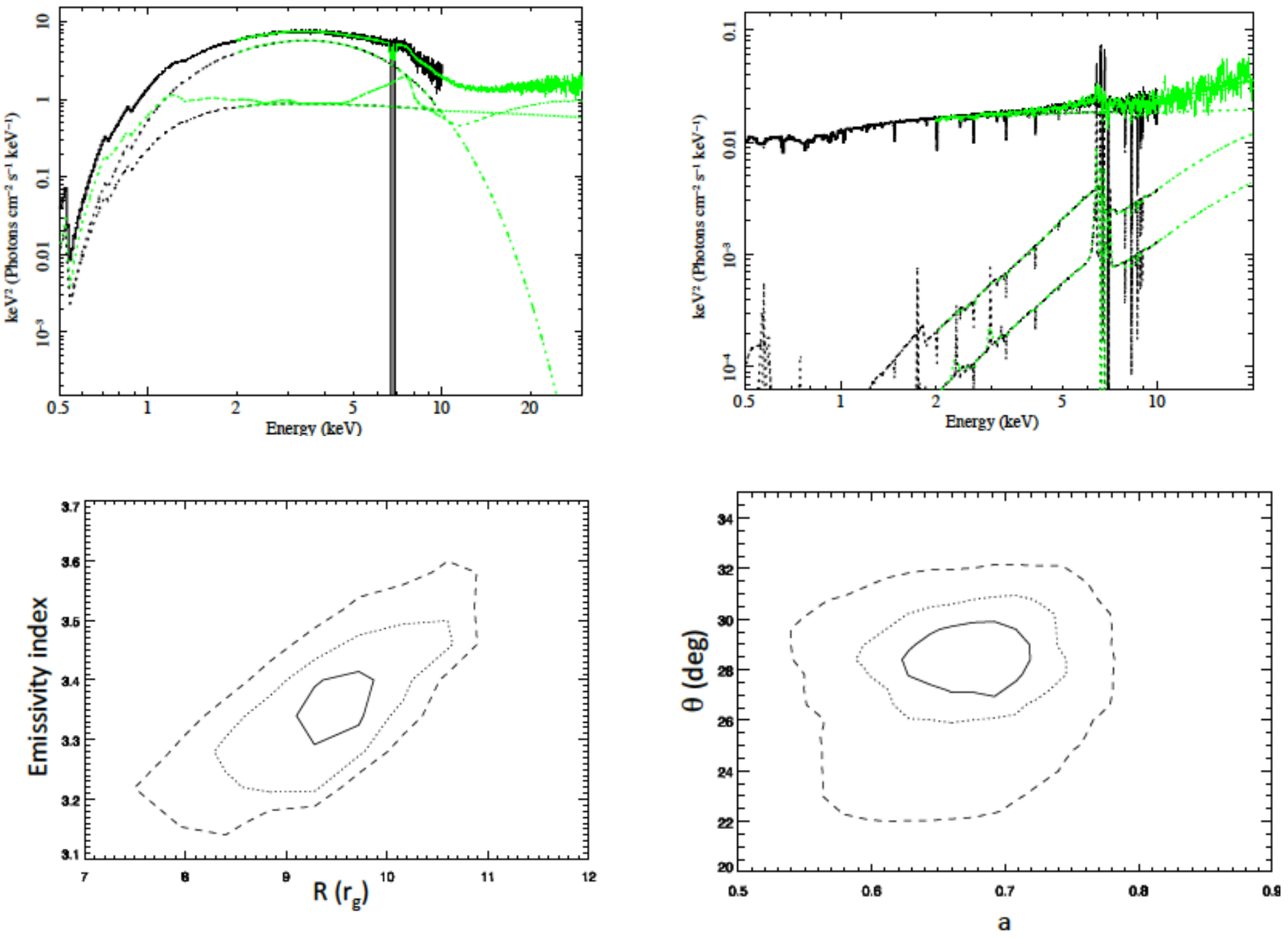

Figure 13. Upper left panel: eXTP broadband (SFA+LDA) spectrum for an exposure of 0.1 ks of a 0.5 Crab, maximally rotating XRB BH such as GRO J1655-40 [46]. Upper right panel: eXTP spectrum obtained with a 100 ks integration of a 2 mCrab, spin a= 0.5 AGN. Lower panels, confidence levels for the disc emission vs. inner radius for XRB (left ) and inclination vs. black hole spin for AGN (right ) (1, 2, and 3 sigma solid, dotted, and dashed lines respectively). *(Courtesy of the eXTP working group on Strong Field Gravity).*

The broad band pass from 0.5 to 30 keV of eXTP, the good energy resolution, and its large photon throughput, enables the reverberation lags of the disk blackbody, Fe K lines and Compton reflection humps to be measured simultaneously, so that each provides a separate measure of the disk inner radius (see Uttley et al. 2014 for a detailed review on reverberation mapping [47]). This allows an accurate and independent determination of the BH spin, if the disk extends to the ISCO. Many more details on the dramatic improvements in reverberation studies especially for XRB, are to be found in the eXTP white paper on accretion in strong-field gravity. Another technique to measure the radius (at ISCO) of the of the accretion disk in XRBs and hence the spin of the BH is given by the continuum-fitting method ([48], [49]). A wide range of spins has been measured with this technique, but in many case errors are large and systematic errors cannot be excluded (see [50] and references therein). By combining its powerful spectral-timing approach eXTP will allow a big step in the accuracy of the use of disk thermal emission fitting to map the innermost regions and measure BH spins.

As a second enlightening example, we show in Fig. 14 the sensitivity (in fractional rms) of the eXTP instruments (LAD, SFA and the combination of the two) for the detection of a QPO with a FWHM of 10 Hz. The left panel shows the case if we fix the exposure time to 10 ks and for a flux variable between 0.001 to 1 Crab (the flux range of most XRBs). The right panel shows the case for a source at 1 Crab flux, and for a variable exposure time between 100 s and 10 ks. eXTP will bring an improvement of at least one order of magnitude in sensitivity compared to RXTE/PCA, still the best instrument allowing high time resolution studies so far. eXTP's large effective area will allow us to measure the QPO waveforms either directly, for QPOs that will for the first time be detected coherently, or by Fourier reconstruction. Coherent detection requires the collection of a sufficient number of photons for detection within the signal's coherence time, and hence can be confidently predicted for signals that are only incoherently detected in current data. So far, coherent detection has been limited to a few, high coherence low frequency QPOs and even in those cases S/N, spectral resolution, or both were insufficient to study the phase dependence of the spectral shape. With eXTP, coherent detection will be common for low-frequency QPOs. Most neutron star kHz QPOs, well detected in previous missions, will be detected for the first time coherently with eXTP. In addition, the LAD will enable phase-resolved spectroscopy of QPOs. With a 50 ks eXTP observation of GRS 1915+105, phase-resolved spectra in 20 QPO phases can be significantly constrained. This implies that the change in shape of the iron line as a function of QPO phase resulting from Lense-Thirring precession of the inner flow can be clearly detected.

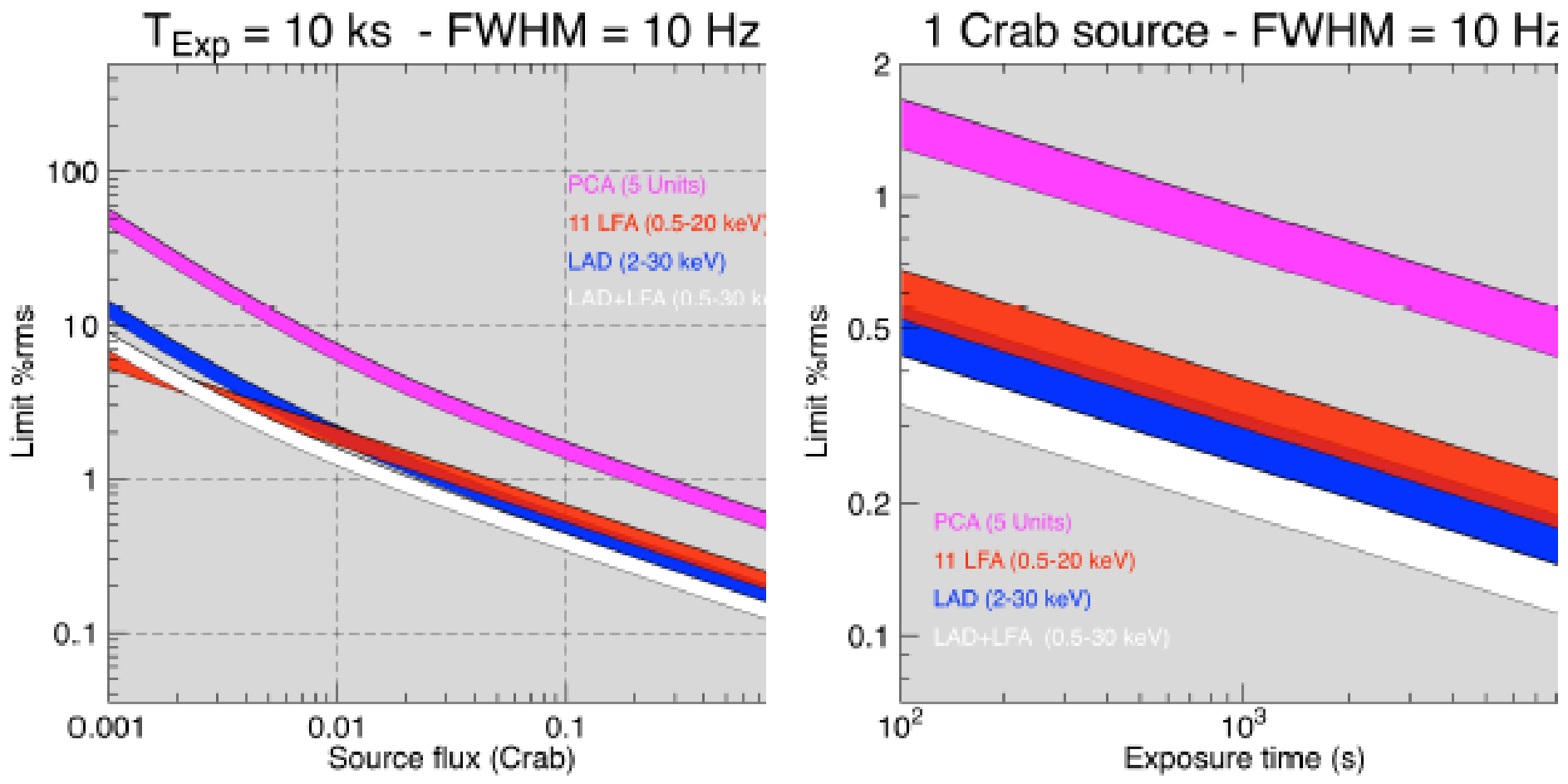

Figure 14. Right Panel: simulated eXTP sensitivity in fractional rms% for QPO with FWHM=10 Hz as a function of exposure for a source with flux equal to 1 Crab. Right Panel: simulated eXTP sensitivity in fractional rms% for QPO with FWHM=10 Hz as a function of flux for 10ks exposure. In both panels each stripe marks the 3 sigma and 5 sigma significance levels (lower and upper boundary, respectively). Different colors correspond to different instrument: RXTE, 5 PCUs (magenta); SFA (erroneously labeled as LFA in the figure), 11 telescopes (red); LAD, 40 elements (blue), sum of SFA (0.5-20 keV) and LAD (0.5-30 keV) (white). *(Courtesy of the eXTP working group on Strong Field Gravity).*

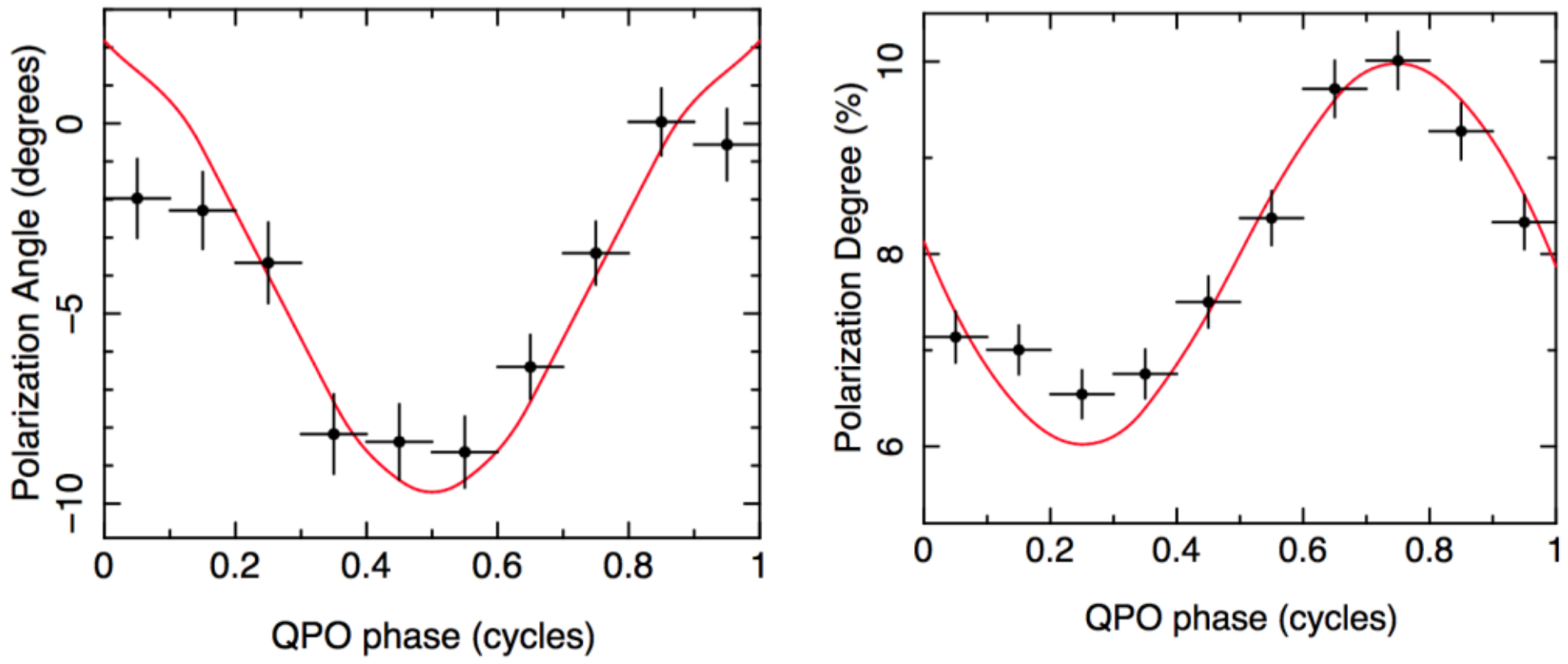

Figure 15. Measured (black points) and input (red lines) polarization degree (left) and angle (right) as a function of QPO phase from our simulation of a 1 Hz QPO. We see that the modulations in polarization properties are recovered well by eXTP . A 1 kHz QPO was simulated for a GRS 1915+105-like source in its intermediate state based and adapted from the RXTE observation reported in [51].

A unique capability of eXTP will be the combination of X-ray polarimetric and timing information. This is enabled by the PFA and the LAD combination, which allows the photon polarization signal obtained by the PFA to be cross-correlated with the very high S/N light curves obtained by the LAD, to allow the time-dependent polarization signal to be extracted using similar approaches to spectral-timing measurements such as reverberation and QPO phase resolved studies. Although the details of the technique are out of the scope of this paper, we wish to show in Figure 15 (black points) the resulting phase-folded polarization degree (left) and angle (right) plotted as a function of QPO phase for a GRS 1915+105-like source. The points are clearly not consistent with constant polarization properties. The red line on each plot shows the input modulation. The technique provides again a powerful independent diagnostics of the accretion flow geometry, and constraints on key parameters like the BH spin. As discussed above, information on the accretion flow and measurements of the BH spin can be independently obtained by iron lines studies, including reverberation mapping and QPO phase resolved spectroscopy, continuum-fitting method and QPO phase resolved polarization studies. In addition, a fourth probe can be used: the energy dependence of the polarization angle and degree of the accretion disk emission, which significantly changes with the BH spin. Such an impressive combination of independent methods makes eXTP unique for strong-field accretion studies.

### 3.4 eXTP as an observatory

With a uniquely high throughput, good spectral resolution, wide sky coverage, and polarimetry capability, eXTP is a powerful observatory very well suited for a variety of studies complementing the core science objectives. The SFA will provide high sensitivity in a soft but wide bandpass of 0.2 to 15 keV, the PFA will provide X-ray polarimetry capability, the LAD will provide the best timing and spectroscopic studies ever for a wide range of high energy sources brighter than 1 mCrab in the 2 to 30 keV band, and the WFM, with its unprecedented combination of field of view and imaging down to 2 keV, makes eXTP a discovery machine of the variable and X-ray transient sky. The WFM will reveal many new sources for follow-up with the SFA, PFA, LAD and other facilities. The WFM will also be monitoring daily hundreds of sources, to catch unexpected events and provide long-term records of their variability and spectroscopic evolution. We observe that no other All Sky Monitor is currently planned for the 2020s. eXTP will be a unique, powerful X-ray partner for other new large-scale facilities across the spectrum likely available in the 2020s, such as gravitational waves and neutrino experiments, SKA and pathfinders in the radio, LSST and E-ELT in the optical, and CTA at TeV energies (Figure 16). In particular eXTP will be a powerful complement for the exploration of the gravitational waves sky, since it will reveal the electromagnetic signals and hence the counterparts associated to many of the still to be discovered sources of gravitational waves. In addition the eXTP's core program will synergically explore key physics issues at the core of gravitational wave astronomy. A number of key targets for the observatory science program (e.g., low-mass X-ray binaries or 'LMXBs') coincide with those that will be observed as part of the eXTP 's core program. Some observatory science goals will thus be pursued from the same observations and do not require additional exposure time. Other targets of the observatory science program (e.g., accreting white dwarfs, blazars, high mass X-ray binaries), can in turn provide useful comparative insights for the core science objectives.

## 4. CONCLUSIONS

Building on the heritage of the XTP study in China and of the LOFT studies in Europe, an international consortium led by the IHEP of the Chinese Academy of Sciences is currently developing the eXTP mission. Thanks to the combination of state-of-the-art instrumentation, eXTP will address fundamental questions of physics and astrophysics, with broadband, high throughput, good energy resolution and polarimetry observations of bright objects of the X-ray sky. Its core objectives make eXTP fully complementary to ESA's ATHENA mission. eXTP will also be a milestone of the multi-messenger exploration of the Universe in the next decade. As of today, the Consortium aims at a launch date before 2025, which is feasible considering the maturity of the many elements of the mission. Currently, the already existing eXTP working groups are being enlarged to an even wider community so to make this challenging mission a truly world-wide effort of the astronomical and astrophysical community.

## ACKNOWLEDGEMENTS

The Chinese team acknowledges the support of the Chinese Academy of Sciences through the Strategic Priority Research Program of the Chinese Academy of Sciences, Grant No. XDA04060600. Authors from Italian institutions acknowledge support by ASI, INFN and INAF. The work of the MSSL-UCL has been supported by the UK Space

Agency. The University of Geneva team at ISDC and DPNC acknowledges the support of the Swiss State Secretariat for Education, Research and Innovation SERI and ESA's PRODEX programme. The Polish team acknowledges Polish National Science Centre grant 2013/10/M/ST9/00729. Czech team acknowledges GA CR grant 13-33324S. The German teams acknowledge the support of the German Space Agency- DLR and of the Max-Planck-Gesellschaft.

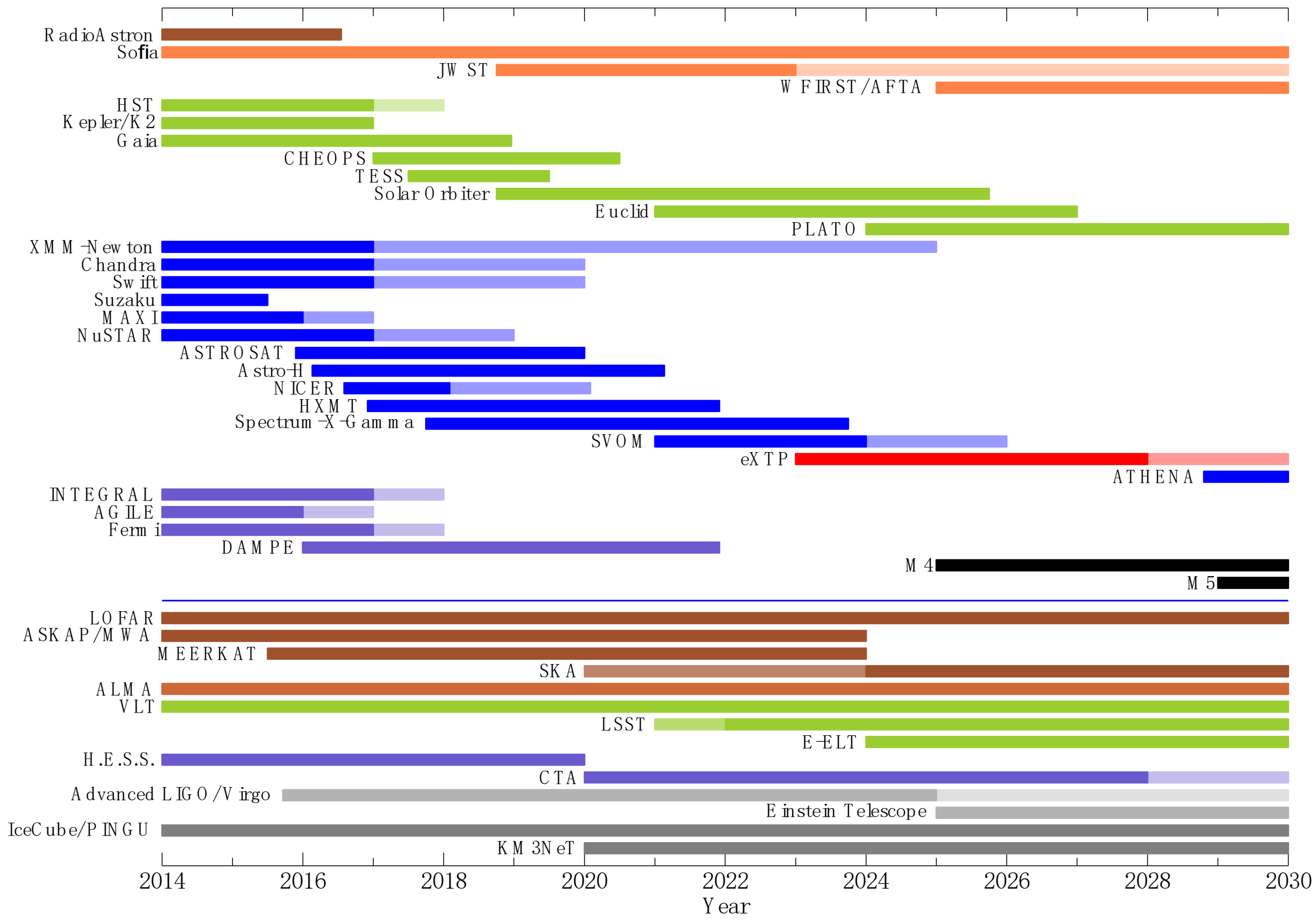

Figure 16. Multi-messenger facilities relevant to eXTP. Colors indicate similar wavebands from the radio (brown) via IR (red) and optical (green) to X-rays (blue) and gamma rays (purple). Grey bands: gravitational wave and neutrino detectors. Dark colors: current end of funding, light colors: expected lifetime, where known, independent of funding decisions. The thin line separates space-based (top) from ground-based (bottom) facilities. *(Courtesy of the eXTP working group on observatory science)*.